\documentclass{epl}
\title{A model for the generic alpha relaxation of viscous liquids}
\author{Jeppe C. Dyre}
\institute{Department of Mathematics and Physics (IMFUFA), ``Glass and time'' - Danish National Research Foundation Centre for Viscous Liquid Dynamics, Roskilde University, Postbox 260, DK-4000 Roskilde, Denmark}
\usepackage{bm}
\begin{document}

\newcommand{\la}{\left\langle}
\newcommand{\ra}{\right\rangle}
\newcommand{\rk}[1]{{\rho_{\bf  {#1}}}}
\newcommand{\nk}[1]{{{\bf n}_{\bf  {#1}}}}
\newcommand{\bk}{{\bf k }}
\newcommand{\bn}{{\bf n}}
\newcommand{\br}{{\bf r}}
\newcommand{\gr}{\Gamma_{\rho}}
\newcommand{\gn}{\Gamma_{{\bf n}}}
\newcommand{\sti}{\tilde s}
\newcommand{\tti}{\tilde t}

\maketitle
\begin{abstract} 
Dielectric measurements on molecular liquids just above the glass transition indicate that alpha relaxation is characterized by a generic high-frequency loss varying as $\omega^{-1/2}$, whereas deviations from this come from one or more low-lying beta processes [Olsen et al, Phys. Rev. Lett. {\bf 86} (2001) 1271]. Assuming that   long-wavelength fluctuations dominate the dynamics, a model for the dielectric alpha relaxation based on the simplest coupling between the density and dipole density fields is proposed here. The model, which is solved in second order perturbation theory in the Gaussian approximation, reproduces the generic features of alpha relaxation. 
\end{abstract}

\section{Introduction}

Alpha relaxation is the slowest and the dominant relaxation process in viscous liquids, closely linked to the glass transition because it sets the time scale for equilibration. Two of the most important unsolved problems regarding viscous liquids concern properties of the alpha process.  One problem \cite{rev} is to explain the temperature dependence of the alpha relaxation time $\tau$, which for some liquids increases by more than a factor of 10 when temperature is lowered by just 1\%. Here we address the second problem: how can the observed shape of the alpha loss peak be explained? 

It was recognized already more than 40 years ago that the dielectric alpha process is always asymmetric, with almost Debye behavior at frequencies lower than the loss-peak frequency whereas there is a non-trivial power law loss, $\epsilon''(\omega)\propto\omega^{-n}$, above the loss peak [$\epsilon(\omega)=\epsilon'(\omega)-i\epsilon''(\omega)$ is the frequency-dependent dielectric constant, $\epsilon''$ the dielectric loss]. In many cases $n$ was found to be close to $1/2$ and over time several theories were proposed predicting $n=1/2$ \cite{oldtheories}. As measurements accumulated apparently contradicting any such universality, it became popular to represent data by the stretched exponential dipole time-autocorrelation function ($\propto\exp[-(t/\tau)^\beta]$ where $t$ is time and $\beta<1$ a parameter). Via the fluctuation-dissipation theorem this function reproduces the Debye behaviour at low frequencies, and it predicts an asymptotic power-law decay above the loss peak with $n=\beta$. Often, however, the high-frequency power law applies only just above the loss peak frequency and a ``wing'' appears at higher frequencies. Nagel and co-workers proposed that, when scaled in an unusual way, the alpha dielectric loss follows a universal curve which includes the wing \cite{nagel}. In 1997 at the 3rd International Discussion Meeting on Relaxations in Complex Systems it was suggested by Olsen, however, that beta processes may play a role at much lower frequencies than previously expected and that this could explain the wing \cite{ols98}. This was confirmed in experiments by Lunkenheimer and co-workers and subsequently by other groups, showing that the wing indeed develops into a separate relaxation process after long-time annealing right below the glass transition \cite{lunk}. A publication from 2001 giving data for ten molecular liquids \cite{ols01} presented indications that $n=1/2$ is a universal high-frequency exponent in the following sense: Equilibrium viscous liquids approach this behaviour as temperature is lowered and effects of beta processes gradually become negligible; in some cases, however, this may happen only when the alpha relaxation time is months or longer, making it impossible to verify or falsify the conjecture that $n=1/2$ is universal.

It is not generally accepted that $n=1/2$ is a generic high-frequency exponent of alpha relaxation. Nevertheless, it makes sense to look for simple theories which can reproduce the proposed generic behaviour; the exponent $n=1/2$ ought to be easier to explain than any other non-trivial exponent. For reasons not to be detailed here, the old models all have problems. In a recent paper it was suggested that a long-time-tail mechanism might be at the root of generic alpha relaxation \cite{dyr05}. Two immediate questions arise if this were correct: 1. The mechanism should apply a few decades above the alpha loss peak frequency, i.e., at times {\it shorter} than the alpha relaxation time; 2. the long-time tail of the relevant velocity autocorrelation function must be negative. The latter problem is solved by assuming stochastic dynamics, whereas a solution of the former is suggested by the solidity of viscous liquids \cite{dyr05}. Reference \cite{dyr05} did not discuss a specific model for the alpha relaxation. This is done below, where a model implementing the long-time-tail mechanism is proposed and solved for $\epsilon(\omega)$ in the simplest approximation.

\section{Model}

Consider $N$ molecules in volume $V$. If $\bn$ is the normalized molecular dipole vector, the basic degrees of freedom are taken to be the density field $\rho(\br)$ and the dipole-density field $\bn(\br)$. It is convenient to transform into k-space by defining the following sums over all molecules

\begin{eqnarray}\label{1}
\rk\bk\ &=&\ \frac{1}{\sqrt N}\sum_j e^{i\bk\cdot\br_ j}\nonumber\\
\nk\bk\ &=&\ \frac{1}{\sqrt N}\sum_j \bn_ j\, e^{i\bk\cdot\br_ j}\,.
\end{eqnarray}
If $\beta$ is the inverse temperature, the Hamiltonian (free energy) for a Gaussian model $H_0$ is given by 
$\beta H_0 =\sum_{\bk} \rk\bk \rk {-\bk}/2S(\bk)+\sum_{\bk} {\nk\bk}\cdot {\nk {-\bk}}/2\epsilon(\bk)$ where $S(\bk)$ is the static structure factor and $\epsilon(\bk)$ the dimensionless wavevector-dependent static dielectric constant. The simplest interaction term is $H_{\rm int}\propto\int \bn^2(\br)\rho(\br)d\br$. We shall assume that the dynamics are dominated by the long-wavelength behaviour of the fields. Denoting the $\bk\rightarrow 0$ (bare) limits of $S(\bk)$ and $\epsilon(\bk)$ by $A$ and $B$ respectively, in terms of a dimensionless coupling constant $\lambda$ the Hamiltonian becomes

\begin{equation}\label{2}
\beta H\ =\
\frac{1}{2A}\sum_{\bk} \rk\bk \rk {-\bk}\,+\,\frac{1}{2B}\sum_{\bk} \nk\bk\cdot \nk {-\bk}\,+\,
\frac{\lambda}{\sqrt N}\sum_{\bk,\bk'}\bn_{\bk}\cdot\bn_{\bk'}\rho_{-\bk-\bk'}\,.
\end{equation}
This defines an ``ultra-local'' field theory, i.e., one where equal-time fluctuations are uncorrelated in space. A more realistic Hamiltonian has gradient terms, but ignoring these is consistent with the assumption of long-wavelength dominance. Since $\rk{\bf 0}$ is always equal to $\sqrt N$, it is not a dynamic degree of freedom and all sums are to be understood excluding terms with $\rk{\bf 0}$. 

Langevin equations define the dynamics:

\begin{eqnarray}\label{3}
\dot{\rho}_{\bk}   &=& -\gr(k)  \frac{\partial (\beta H)}{\partial \rho_{-\bk}}\,+\,\xi_\bk(t)  \nonumber\\
\dot{\bn}_{\bk} &=& -\gn(k)\frac{\partial (\beta H)}{\partial \bn_{-\bk}} \,+\,{\bm\eta}_\bk(t)\,,
\end{eqnarray}
where $\xi_\bk(t)$ and ${\bm\eta}_\bk(t)$ are standard Gaussian white noise terms. Because density is conserved one expects $\gr(k)\propto k^{2}$. We define $\gr(k)/A\equiv Dk^{2}$. The dipole-density field, on the other hand, is not conserved. Consistent with long-wavelength dominance we identify $\gn(k)$ with its $k\rightarrow 0$ limit by writing $\gn(k)/B\equiv\Gamma$. Substituting eq.~(\ref{2}) into eq.~(\ref{3}) now leads to 

\begin{eqnarray}\label{4}
\dot{\rho}_{\bk}   &=& - D k^2 \left(\rho_{\bk}\,+\,\frac{\lambda A}{\sqrt N}\sum_{\bk'}\nk{\bk+\bk'}\cdot\nk{-\bk'}\right)\,+\,\xi_\bk(t)
\nonumber\\
\dot{\bn}_{\bk} &=& -\rm\Gamma \left(\nk\bk\,+\,\frac{2\lambda B}{\sqrt N}\sum_{\bk'}\nk{\bk+\bk'}\rk{-\bk'}\right) \,+\,{\bm\eta}_\bk(t)\,\,.
\end{eqnarray}
Equations (\ref{2}) and (\ref{4}) imply that to zero'th order in $\lambda$ the autocorrelation functions are given by

\begin{eqnarray}\label{5}
&\la\rk\bk(0)\rk{-\bk}(t)\ra_0 &\,=\,\,A\,e^{-Dk^2t}\nonumber\\
&\la\nk\bk(0)\cdot\nk{-\bk} (t)\ra_0 &\,=\,\,3B\,e^{-\Gamma t}\,.
\end{eqnarray}
We proceed to calculate the dipole autocorrelation function by applying the following theorem: If $Q_i$ obeys a Langevin equation of the form $\dot Q_i=-\mu \partial_i H + \xi_i(t)$, one has (no sum over $i$) $d^2/dt^2\la Q_i(0)Q_i (t)\ra=\mu^2\la \partial_i H(0)\partial_i H(t)\ra$ \cite{dyr05,ris89}. To second order in $\lambda$ the dipole autocorrelation function in the Gaussian approximation thus obeys 

\begin{eqnarray}\label{6}
\Gamma^{-2}\frac{d^2}{dt^2}\la\nk\bk(0)\cdot\nk{-\bk}(t)\ra\,&=&\,
\la\nk\bk(\rm 0)\cdot\nk{-\bk}(t)\ra\,\nonumber\\
&+&\,\frac{4\lambda^2 B^2}{N}\sum_{\bk'}\la\nk{\bk+\bk'}(0)\cdot\nk{-\bk-\bk'}(t)\ra_0\la\rk{-\bk'}(0)\rk{\bk'}(t)\ra_0\,.
\end{eqnarray}
The sum is equal to $3ABVe^{-\Gamma t}(Dt)^{-3/2}/32\pi^{5/2}$. In terms of the dimensionless time $\tti\equiv \Gamma t$, if the constant $3\lambda^2 AB^3(\Gamma/D)^{3/2}V/N 8\pi^{5/2}$ is denoted by $\Lambda$, the dipole autocorrelation function $\Phi(\tti)\equiv \la\nk\bk(0)\cdot\nk{-\bk}(\tti)\ra$ is the solution of

\begin{equation}\label{7}
\ddot\Phi\,=\,
\Phi\,+\,\Lambda  e^{-\tti}\tti^{-3/2}\,.
\end{equation}
The general solution to this equation obeying $\Phi(\tti\rightarrow\infty)=0$ is 

\begin{equation}\label{8}
\Phi(\tti)\,=\,
\Lambda\int_{\tti}^\infty \sinh({\tti}'-{\tti})e^{-\tti'}\tti'^{\rm-3/2}d\tti'\,+\,C_2\,e^{-\tti}\,,
\end{equation}
which can be expressed in terms of the error function. In dimensionless units the dielectric constant is the Laplace transform of $-\dot\Phi$, leading (Appendix) to (where $C_1=\Lambda\sqrt\pi$ and $\tau\equiv1/\Gamma$)

\begin{equation}\label{9}
\epsilon(\omega)\,=\,
C_1\left(\frac{1}{\,\sqrt{1+i\omega\tau}\,}+\frac{1}{\,\sqrt 2+\sqrt{1+i\omega\tau}\,}\right)
\,+\,C_2\,\frac{1}{\,1+i\omega\tau\,}\,.
\end{equation}
This function has the required asymptotic behaviours: $\epsilon''(\omega)\propto\omega$ for $\omega\tau\ll 1$ and 
$\epsilon''(\omega)\propto\omega^{-1/2}$ for $\omega\tau\gg 1$. There is one overall scaling parameter and one non-trivial parameter determining the loss-peak shape around the maximum.

\section{Discussion}

It has been shown that generic features of the alpha process are reproduced by a model based on a conserved scalar field and a non-conserved vector field with the simplest possible interaction term. The basic assumption of the model is that the dynamics are dominated by long-wavelength fluctuations. This justifies taking $\Gamma_\rho(k)\propto k^2$ and that $S(k)$ and $\epsilon(k)$ may be replaced by their $k\rightarrow 0$ limits. Note that long-wavelength dominance is unrelated to possibly having a diverging length scale upon cooling. It simply means that the coherent diffusion constant $D$ is much larger than estimated from $\tau$ and the average intermolecular distance $a$ ($D\gg a^2/\tau$) -- whenever this applies, density fluctuations on length scales of order 10$a$-100$a$ take place on the alpha time scale and influence the dielectric loss, even above the loss peak frequency. The resulting dipolar dynamics have no length-scale dependence.

An argument for long-wavelength dominance of the dynamics is given in a recent publication \cite{dyr05}, where it is argued that the solidity of  highly viscous liquids implies that the coherent diffusion constant $D$ is much larger than the incoherent diffusion constant $D_s$ [defined from the long-time mean-square displacement of one molecule]. The reasoning, which assumes that the coherent diffusion constant is frequency independent, is briefly summarized as follows. At high viscosity molecular motion occurs via ``inherent dynamics'', i.e., jumps between potential energy minima \cite{inhdyn}. Momentum conservation is irrelevant when relaxation times are in the second range or longer (the liquid container walls contribute any required momentum). Thus if $X$ is the sum of all molecular x-coordinates, it makes sense to define a ``center of mass diffusion constant'' by $D_{\rm CM}=\lim_{t\rightarrow\infty}\lim_{V\rightarrow\infty}\la \Delta X^2(t)\ra/2Nt$, where $N/V=$ constant as $V\rightarrow\infty$. For the high-frequency limit $D_{\rm CM}(\infty)$ one finds $S(0)=D_{\rm CM}(\infty)/D$ which, if $D$ is assumed to be frequency independent, implies $D_{\rm CM}(\infty)\ll D$. Moreover, it is argued that $D_s(\infty)\sim D_{\rm CM}(\infty)$ and $D_s(\infty)\gg D_s$, where $D_s$ is the dc limit of $D_s(\omega)$. Naively one expects $D_s\sim a^2/\tau$, but the frequently observed breakdown of the Debye-Stokes-Einstein relation \cite{breakdown} implies that $D_s\gg a^2/\tau$. In either case, combining all these inequalities leads to $D\gg a^2/\tau$. If one assumes that $\Gamma_\rho(k)\propto k^2$ applies when $1/\Gamma_\rho$ is longer than the characteristic time $t_c\equiv a^2/D$, we find that, since $t_c\ll \tau$, $\Gamma_\rho(k)\propto k^2$ applies for a range of rates larger than the loss peak frequency. -- Note that, even if diffusion constant decoupling does not apply and $D\sim D_s$, long-wavelengths dominate the dynamics whenever the Debye-Stokes-Einstein relation is violated.

The model only applies for extremely viscous liquids (with viscosity at least $10^{10}$ times larger than that of ambient water), because only in this limit do liquids have the solidity property. Experimentally, this is also where the $\omega^{-1/2}$ high-frequency behaviour is asymptotically approached. We conjecture that in the extremely high viscosity limit (but still in the equilibrium liquid phase) an $\omega^{-1/2}$ high-frequency universality always appears. This, in fact,  should apply to any linear response function. 

A few final comments:

1. The assumption that $\Gamma_\rho(k)\propto k^2$ cannot -- despite its hydrodynamic flavour -- apply all the way down to $k=0$. At very large wavelengths a k-independent density ``decay channel'' must appear, because otherwise there is no bulk volume relaxation on the alpha time scale and the glass would have the same compressibility as the equilibrium liquid. Thus a more accurate model would assume $\Gamma_\rho(k)=\Gamma_0+D k^2$ (which, however, only slightly perturbs model predictions if $\Gamma_0\sim1/\tau$ as required by experiment).

2. The Hamiltonian of eq.~(\ref{2}) is ``bottomless'' in the sense that it allows arbitrarily low energies \cite{gre84}, implying that the canonical distribution is not normalizable. This inconsistency is readily resolved by assuming that there are higher even order terms in ${\bf n}(\br)$. If these terms are small, they barely influence the equilibrium autocorrelation functions, but still ensure that the fields do not eventually run off to infinity.

3. According to the model the generic $\omega^{-1/2}$ high-frequency decay of the alpha process is specific to three dimensions. In four dimensions, for instance, the sum in eq.~(\ref{6}) varies with time as $t^{-2}$, leading to an autocorrelation function which varies logarithmically with time as $t\rightarrow 0$.

--------------------------------------------------------

H B Nielsen, G Ruocco and T Schr{\o}der are thanked for helpful comments. This work was supported by the Danish National Research Foundation.

\acknowledgments

APPENDIX

In order to calculate the Laplace transform of $-\dot\Phi$, we first note that the second term of eq. (\ref{8}) leads to the $C_2$-term of eq. (\ref{9}). The first term requires calculating the following integral (where $s\equiv i\omega$ and tildes are left out for simplicity of notation):

\[ 
I =\int_0^\infty dt\ e^{-st} \int_t^\infty\cosh(t'-t)e^{-t'}t'^{-3/2}dt'=\int_0^\infty dt\ e^{-st} \int_t^\infty \frac{e^{t'-t}+e^{t-t'}}{2}e^{-t'}t'^{-3/2}dt'.
 \]
We split $I$ into three terms:

\[ 
I =\frac{1}{2}\int_0^\infty dt\ e^{-(s+1)t} \int_t^\infty t'^{-3/2}dt'
+\frac{1}{2}\int_0^\infty dt\ e^{-(s-1)t} \int_t^\infty t'^{-3/2}dt'+(*),
 \]
where

\[
(*)=\frac{1}{2}\int_0^\infty dt\ e^{-(s-1)t} \int_t^\infty (e^{-2t'}-1)t'^{-3/2}dt'.
\]
Thus

\[
I=\int_0^\infty \ e^{-(s+1)t} t^{-1/2}dt+\int_0^\infty \ e^{-(s-1)t} t^{-1/2}dt+(*).
\]
Since $\int_0^\infty e^{-t}t^{-1/2}dt=\sqrt\pi$ we get

\[
I=\sqrt\pi(1+s)^{-1/2}+\sqrt\pi(s-1)^{-1/2}+(*).
\]
The last term is calculated as follows:

\[
(*)=\frac{1}{2}\left[\frac{1}{1-s}e^{-(s-1)t}\int_t^\infty (e^{-2t'}-1)t'^{-3/2}dt'\right]_{t=0}^{t=\infty}
-\frac{1}{2}\frac{1}{1-s}\int_0^\infty e^{-(s-1)t}(1-e^{-2t})t^{-3/2}dt.
\]
Thus

\[
(*)=\frac{1}{2}\frac{1}{1-s}\left(\int_0^\infty (1-e^{-2t})t^{-3/2}dt
-\int_0^\infty \left(e^{-(s-1)t}-e^{-(s+1)t}\right)t^{-3/2}dt\right),
\]
or

\[
(*)=\frac{1}{1-s}\Big(F(0,2)-F(s-1,s+1)\Big)
\]
where

\[
F(a,b)=\frac{1}{2}\int_0^\infty \left(e^{-at}-e^{-bt}\right)t^{-3/2}dt.
\]
Calculating this function is straightforward:

\[
F(a,b)=\left[(e^{-at}-e^{-bt})(-t^{-1/2})\right]_{t=0}^{t=\infty}
-\int_0^\infty \left(ae^{-at}-be^{-bt}\right)t^{-1/2}dt
=\phi(b)-\phi(a),
\]
where

\[
\phi(x)=\int_0^\infty xe^{-xt}t^{-1/2}dt=\sqrt x\sqrt\pi.
\]
Summarizing,

\[
\frac{I}{\sqrt\pi}=
(1+s)^{-1/2}+(s-1)^{-1/2}+\frac{1}{1-s}\left(\sqrt 2-\sqrt 0-\sqrt{s+1}+\sqrt{s-1}\right),
\]
or

\[
\frac{I}{\sqrt\pi}=(1+s)^{-1/2}+\frac{\sqrt 2 - \sqrt{1+s}}{1-s}
=\frac{1}{\sqrt{1+s}}+\frac{1}{\sqrt 2 + \sqrt{1+s}}.
\]

\end{document}